\documentclass[12pt,showpacs,showkeys,amsmath,amssymb]{revtex4}
\usepackage{amsmath,amsfonts,amsthm,amscd,amssymb,latexsym}
\usepackage{bm}
\usepackage{dcolumn}
\usepackage{graphicx}
\usepackage{epstopdf}
\usepackage{color}
\usepackage{epsf}
\usepackage{epsfig}
\usepackage{graphicx, epic, eepic, color}

\newcommand{\beq}{\begin{equation}}
\newcommand{\eeq}{\end{equation}}

\begin{document}

\title{Conformal Symmetry, Accelerated Observers and Nonlocality}

\author{Bahram \surname{Mashhoon}$^{1,2}$}
\email{mashhoonb@missouri.edu}

\affiliation{$^1$Department of Physics and Astronomy, University of Missouri, Columbia, Missouri 65211, USA\\
$^2$School of Astronomy, Institute for Research in Fundamental
Sciences (IPM), P. O. Box 19395-5531, Tehran, Iran\\
}

\date{\today}

\begin{abstract}
The acceleration transformations form a 4-parameter Abelian subgroup of the conformal group of Minkowski spacetime. The passive interpretation of acceleration transformations leads to a  congruence of uniformly accelerated observers in Minkowski spacetime. The properties of this congruence are studied in order to illustrate the kinematics of accelerated observers in relativistic physics. The generalization of this approach under conformal rescaling of the spacetime metric is examined. 
\end{abstract}


\maketitle

\section{Introduction}

 In the absence of gravitation, Minkowski spacetime provides the flat background for special-relativistic physics. The observers in this arena are either hypothetical inertial observers or accelerated observers. Inertial observers, each forever spatially at rest in an inertial frame of reference, were introduced into physics by Newton and play an essential role since the fundamental laws of physics have been formulated with respect to these nonexistent observers. Realistic observers are all more or less accelerated. To access the laws of physics, the measurements of accelerated observers must somehow be related to the hypothetical inertial observers. 
 
 The measurements of inertial observers are related to each other via \emph{Lorentz invariance}. What do accelerated observers measure? The standard answer within the framework of the special theory of relativity involves performing Lorentz transformations point by point along the world line of the accelerated observer~\cite{Einstein}. That is, the standard prescription assumes the \emph{locality postulate}, which states that an accelerated observer is pointwise equivalent to an otherwise identical momentarily comoving inertial observer~\cite{Mash1, Mash2}. The locality postulate of relativity theory is strictly valid if all measurements are spatially pointwise and instantaneous. The internal mechanisms of measuring devices carried by accelerated observers could be subject to Coriolis, centrifugal and other inertial effects that may add up to influence the results of measurements. We assume, however, that all such devices are \emph{standard}; that is, they are sufficiently robust against inertial effects and thus function in accordance with the locality postulate. 

These notions of the standard relativity theory are based on the assumption that all physical phenomena could be reduced to \emph{pointlike coincidences}.  However, Bohr and Rosenfeld have pointed out that the measurement of the electromagnetic field cannot be performed instantaneously and generally involves an average over a spacetime domain~\cite{BR1, BR2}. Similarly, wave phenomena are generally nonlocal by the Huygens principle. Consider, for instance, the measurement of the frequency of an incident electromagnetic wave by an accelerated observer. We expect that the locality assumption is a good approximation if the wavelength of the incident radiation $\lambda$ is sufficiently small compared to the length scale $L$ characteristic of the observer's acceleration. In fact, $L/c$ is the effective length of  time over which the state of the accelerated observer changes appreciably.  For observers fixed on the Earth, the corresponding translational acceleration length is about 1 light year and the rotational acceleration length is about 28 astronomical units; therefore, the locality postulate is normally an excellent approximation since $\lambda \ll L$. On the other hand, for observers undergoing large accelerations, $\lambda \gtrsim L$, the past history of the accelerated observer must be taken into account in accordance with nonlocal special relativity~\cite{Mash3a, Mash3b, Mash3c}. Nonlocal special relativity goes beyond the locality postulate by including a certain linear average over the past world line of the accelerated observer~\cite{Mash5, BF}. Furthermore, a classical nonlocal generalization of Einstein's theory of gravitation has been developed in analogy with the nonlocal electrodynamics of media. In this theory, nonlocal gravity simulates dark matter~\cite{HM1, HM2, Mash3c}. 

The main purpose of this paper is to describe the kinematics of a congruence of accelerated observers related to the special conformal transformations (Section II). This particular  uniformly accelerated system is then employed in Section III to illustrate the main aspects of accelerated observers in Minkowski spacetime. The extensions of our treatment under conformal rescaling of the spacetime metric  are explored in Section IV. Section V contains a brief discussion of our results. We use units such that $c = 1$, unless specified otherwise; moreover, the signature of the spacetime metric is +2 and greek indices run from 0 to 3, while latin indices run from 1 to 3.

\section{Conformal Symmetry of Minkowski Spacetime}

The conformal group of Minkowski spacetime consists of all coordinate transformations that leave the light cone invariant~\cite{B1, E1, B2, FRW, KA}. This 15-parameter Lie group includes the 10-parameter Poincar\'e group, the 1-parameter scale transformation $x^\mu \mapsto \sigma \,x^\mu$, where $\sigma$ is a constant, and the 4-parameter acceleration transformation $x^\mu \mapsto x'^\mu$, 
\begin{equation}\label{U1}
 x'^\mu = \frac{x^\mu + a^\mu\,x^2}{1+2\,a\cdot x + a^2\,x^2}\,,
\end{equation}
where $a^\mu$, $\mu = 0, 1, 2, 3$, are constant acceleration parameters, each of dimensions 1/length. Here, $a^2 := \eta_{\alpha \beta}\, a^\alpha a^\beta$, $x^2 := \eta_{\alpha \beta}\, x^\alpha x^\beta$ and $a \cdot x := \eta_{\alpha \beta}\, a^\alpha x^\beta$. The coordinate transformation~\eqref{U1}, which leaves the origin of spacetime coordinates invariant, is admissible if 
\begin{equation}\label{U2}
 D(x) = 1+2\,a\cdot x + a^2\,x^2 \ne 0\,.
\end{equation}
We must therefore exclude from transformation~\eqref{U1} events $x^\mu$ for which $D(x) = 0$. If $a^2 \ne 0$, then
\begin{equation}\label{U3}
 \frac{1}{a^2} \,D(x) =  \eta_{\alpha \beta}\, \left(x^\alpha +  \frac{a^\alpha}{a^2}\right)\left(x^\beta +  \frac{a^\beta}{a^2}\right)\,.
\end{equation}
Thus we must exclude from transformation~\eqref{U1} all events $x^\alpha$ on the null cone centered at $- a^\alpha / a^2$. On the other hand, if $a^2 = 0$, then all events on the null hyperplane $1+2\,a\cdot x =0$ are excluded. 

Let us note that when the special conformal transformation~\eqref{U1} is admissible, 
\begin{equation}\label{U4}
x'^2 = \frac{x^2}{1+2\,a\cdot x + a^2\,x^2}\,, 
\end{equation}
so that when $x^2 = 0$, $x'^2 = 0$ as well; hence, the light cone remains invariant. Moreover, when $x^2 \ne 0$ and $x'^2 \ne 0$, Eqs.~\eqref{U1} and~\eqref{U4} imply
\begin{equation}\label{U5}
\frac{x'^\mu}{x'^2} = \frac{x^\mu}{x^2} + a^\mu\,,
\end{equation}
where $x^\mu \mapsto x^\mu / x^2$ is an inversion. Denoting an admissible acceleration transformation by $\mathbb{C}(a^\mu)$, we find $\mathbb{C}(a^\mu)\,\mathbb{C}(b^\mu) = \mathbb{C}(a^\mu+ b^\mu)$; moreover, $\mathbb{C}^{-1}(a^\mu) = \mathbb{C}(-a^\mu)$ and $\mathbb{C}(0) = \mathbb{I}$, where $\mathbb{I}$ is the identity transformation. It follows from these observations that the acceleration transformations form an Abelian subgroup of the conformal group of Minkowski spacetime. 

\section{Accelerated System}

We now adopt a strictly \emph{passive} interpretation of acceleration transformation~\eqref{U1} and note that the Minkowski spacetime interval $ds^2 = \eta_{\mu \nu}\,dx^\mu\,dx^\nu$  can be written in the new coordinates as~\cite{Mash4}
\begin{equation}\label{U6}
ds^2 = g'_{\alpha \beta}(x')\,dx'^\alpha dx'^\beta\,, \qquad   g'_{\alpha \beta} = \frac{\eta_{\alpha \beta}}{(1-2\,a\cdot x' + a^2\,x'^2)^{2}}\,,
\end{equation}
where
\begin{equation}\label{U7}
(1+2\,a\cdot x + a^2\,x^2)(1-2\,a\cdot x' + a^2\,x'^2) = 1\,.
\end{equation}
With an admissible coordinate transformation~\eqref{U1}, the curvilinear coordinates in metric~\eqref{U6} are admissible. To investigate the nature of the accelerated system in Minkowski spacetime, we consider the congruence of \emph{spatially static} observers in the new curvilinear coordinate system. Each observer remains spatially at rest in the accelerated system and carries along its world line a  tetrad frame $e'^{\mu}{}_{\hat \alpha}$ that is  \emph{orthonormal}, namely, 
\begin{equation}\label{U7a}
g'_{\mu \nu}(x')\,e'^{\mu}{}_{\hat \alpha}(x')\,e'^{\nu}{}_{\hat \beta}(x') = \eta_{\hat \alpha \hat \beta}\,.
\end{equation}
Here, the hatted tetrad indices specify the tetrad axes in the tangent space at event $x'$. For the static observers under consideration here, it is natural to choose the tetrad frame such that
\begin{equation}\label{U8}
e'^{\mu}{}_{\hat \alpha} = f(x') \,\delta^\mu_\alpha\,, \qquad f(x') = 1-2\,a\cdot x' + a^2\,x'^2\,.
\end{equation} 
In particular, the world line of such a static observer can be determined via
\begin{equation}\label{U9}
\frac{dx'^0}{d\tau} = f(x')\,, \qquad \frac{dx'^i}{d\tau} = 0\,,
\end{equation} 
where $\tau$ is the proper time along the path of the observer. If $a^2 \ne 0$, we find
\begin{equation}\label{U10}
x'^0(\tau) = \frac{a^0}{a^2} + p \tanh(a^2 \,p\,\tau - q)\,,
\end{equation} 
where $p>0$ is a constant given by
\begin{equation}\label{U11}
p^2 := \delta_{ij}\,\left(x'^i -  \frac{a^i}{a^2}\right)\left(x'^j - \frac{a^j}{a^2}\right)\,
\end{equation} 
and $q$ is an integration constant. In fact, $\tanh{q} = a^0/(a^2\,p)$ once we assume that $x'^0 = 0$ at $\tau = 0$. In the special case that $p =0$, $x'^i = a^i / a^2$  and $a^0 \ne 0$, we get instead of Eq.~\eqref{U10}, 
\begin{equation}\label{U12}
x'^0(\tau) = \frac{a^0}{a^2}\,\frac{a^0\,\tau}{a^0\,\tau -1}\,,
\end{equation} 
so that as $\tau$ approaches $1/a^0$, $x'^0$ diverges. Furthermore, if $a^2 = 0$, then $a^0 \ne 0$ and the temporal coordinate of the accelerated observer varies with its proper time as
\begin{equation}\label{U13}
x'^0(\tau) = \frac{1}{2a^0}( 1 - 2 \delta_{ij}\,a^i\,x'^j) \left(e^{2a^0 \tau} - 1 \right)\,.
\end{equation} 

Let us next turn to the translational acceleration of the static observer, $A'^\mu$, given by
\begin{equation}\label{U14}
A'^\mu = \frac{D'}{d\tau}\,e'^{\mu}{}_{\hat 0}\,,
\end{equation} 
where the covariant differentiation here involves the Christoffel symbols for metric~\eqref{U6} in curvilinear coordinates, namely, 
\begin{equation}\label{U15}
\Gamma'^{\alpha}_{\beta \gamma} = -\frac{1}{f}\,(f_{,\gamma}\,\delta^{\alpha}_{\beta} + f_{,\beta}\,\delta^{\alpha}_{\gamma} - f_{,\delta}\,\eta^{\delta \alpha}\eta_{\beta \gamma})\,.
\end{equation} 
In our convention,  a comma denotes partial differentiation. It is then straightforward to calculate the acceleration 4-vector using Eqs.~\eqref{U14}--\eqref{U15}; the result is
\begin{equation}\label{U16}
A'^\mu = - \eta^{\mu i}\, f\,f_{,i}\,,
\end{equation} 
so that $A'^0 = 0$ and $A'^i = 2(a^i - a^2 x'^i)f(x')$, for $i = 1,2,3$. The translational acceleration 4-vector of the observer as measured by the static observer itself is constant and is given by
\begin{equation}\label{U17}
A'_{\hat \mu} = A'_\alpha\,e'^{\alpha}{}_{\hat \mu} = - \delta_\mu^i\,f_{,i}\,.
\end{equation} 
The constant magnitude of the static observer's acceleration is thus $2\,p\,|a^2|$, since
\begin{equation}\label{U18}
A'_{\hat \mu}\,A'^{\hat \mu} = 4\,a^4\,p^2\,
\end{equation}
and $p^2$ is defined in Eq.~\eqref{U11}.  From the speed of light and the magnitude of the acceleration, one can construct the translational acceleration length of an observer in this congruence, namely, 
\begin{equation}\label{U18A}
L = \frac{c^2}{2\,p\,|a^2|}\,.
\end{equation}

Let us briefly digress here and examine the nature of the uniformly accelerated system under consideration. For $a^2 \ne 0$, each observer that is  spatially at rest in curvilinear coordinates is in general  uniformly accelerated, but the \emph{magnitude of acceleration is different for different observers}. There is an exception, however. The static observer with $x'^i = a^i/a^2$ follows a geodesic and has zero translational acceleration; indeed, this observer is \emph{inertial} and its inertial motion can be determined from Eq.~\eqref{U1} via its inverse transformation, namely, 
\begin{equation}\label{U18a}
x^i = \beta^i \,x^0 + b^i\,, \qquad \beta^i := \frac{2 a^0 a^i}{a^2 + 2\,{a^0}^2}\,, \qquad b^i := -  \frac{a^i}{a^2 + 2\,{a^0}^2}\,.
\end{equation}
Similarly, in terms of inertial coordinates, the uniformly accelerated motion of the observer with $x'^\mu = (x'^0, 0, 0, 0)$ that remains at rest at the spatial origin of the curvilinear coordinate system can be expressed as 
\begin{equation}\label{U18b}
x^i = \frac{a^i}{2\,\alpha^2}\,\left(\sqrt{4\,\alpha^2\,{x^0}^2 + 1} - 1\right)\,,  \qquad \alpha^2 := \delta_{ij}\,a^ia^j \ne 0\,.
\end{equation}
On the other hand, if $a^2 = 0$, then all accelerated observers have the \emph{same} uniform acceleration with magnitude $2\,|a^0|$; that is, 
\begin{equation}\label{U18c}
A'_{\hat \mu}\,A'^{\hat \mu} = 4\,{a^0}^2 = 4\,\delta_{ij}\,a^ia^j\,
\end{equation}
and the corresponding translational acceleration length is $L = c^2/ (2\,|a^0|)$. For treatments of uniformly accelerated motion in the theory of relativity, see, for instance, Refs.~\cite{Born, L+L, FS1, FS2, SF1}. 

For measurement purposes, the static observer's spatial frame $e'^{\mu}{}_{\hat i}$, for $i = 1, 2, 3$, is carried along its world line.   It is possible to show explicitly that the spatial frame of the observer is indeed Fermi-Walker transported along its path. That is, each component $e'^{\mu}{}_{\hat i}$ satisfies the equation of Fermi-Walker transport, namely, 
\begin{equation}\label{U19}
\frac{d\mathbb{S}^\mu}{d\tau}+\Gamma'^{\mu}_{\alpha \beta}\, e'^{\alpha}{}_{\hat 0}\,\mathbb{S}^{\beta}= (A' \cdot \mathbb{S})\,e'^{\mu}{}_{\hat 0} - (e'_{\hat 0} \cdot \mathbb{S})\,A'^\mu\,,
\end{equation}
where $\mathbb{S}^\mu$ is Fermi--Walker transported along  $e'^{\mu}{}_{\hat 0}$.

We have thus far demonstrated that observers that are spatially static in the accelerated system under consideration have in general uniform translational accelerations and their spatial frames are locally nonrotating as they propagate along their world lines. Employing the method of moving frames, it is useful at this point to introduce the general acceleration tensor, namely,  
\begin{equation}\label{U20}
\frac{D'}{d\tau}\,e'^{\mu}{}_{\hat \alpha} = \Phi'_{\hat \alpha \hat \beta}\,e'^{\mu \hat \beta}\,,
\end{equation} 
where 
\begin{equation}\label{U21}
\Phi'_{\hat \alpha \hat \beta} = - \Phi'_{\hat \beta \hat \alpha}\,,
\end{equation} 
is the antisymmetric acceleration tensor of the observer with orthonormal tetrad frame $e'^{\mu}{}_{\hat \alpha}$. For the static uniformly accelerated congruence of observers under consideration here, we find
\begin{equation}\label{U22}
\Phi'_{\hat \alpha \hat \beta} = \eta_{\hat \alpha \hat 0}\,f_{,\beta} - \eta_{\hat \beta \hat 0}\,f_{,\alpha}\,.
\end{equation} 
In analogy with the Faraday tensor, the acceleration tensor can be decomposed into its ``electric" and ``magnetic" parts.  The electric part, $\Phi'_{\hat 0 \hat \alpha} = A'_{\hat \alpha}$, represents the invariant translational acceleration of the observer and the magnetic part, $\Phi'_{\hat i \hat j} = \epsilon_{\hat i \hat j \hat k}\,\omega'^{\hat k}$, represents the observer's invariant rotational acceleration. The latter is the proper rate of rotation of the observer's spatial frame with respect to a locally Fermi-Walker transported frame. The  acceleration scales can be constructed from the invariants of the acceleration tensor~\cite{Mash2}. We have demonstrated that the accelerated observers under consideration have uniform translational accelerations and zero rotational accelerations. The generalization of these results is the subject of the next section.

\section{Conformal Invariance}

Two spacetimes with metric tensors $g_{\mu \nu} (x)$ and $\tilde{g}_{\mu \nu} (x)$ are conformally related if
\begin{equation}\label{V1}
\tilde{g}_{\mu \nu} (x) = \Omega^2(x)\,g_{\mu \nu} (x)\,, \qquad \Omega(x) > 0\,.
\end{equation} 
If the coordinate system is admissible in one spacetime, it is also admissible in the conformally related spacetime. Conformal invariance preserves angles and leaves the local light cone invariant, but can change lengths in a pointwise manner. Indeed, the corresponding spacetime intervals are related via $d\tilde{s} = \Omega(x)\,ds$. Moreover, null geodesics are conformally invariant. 

In this section, we go beyond Minkowski spacetime and consider accelerated observers in a gravitational field. Imagine an accelerated observer in a spacetime with metric $ds^2 = g_{\mu \nu} (x)\,dx^\mu \,dx^\nu$ following a world line with proper time $\tau$. The observer carries an orthonormal tetrad frame $e^{\mu}{}_{\hat \alpha}$ such that its acceleration tensor is given by
\begin{equation}\label{V2}
\frac{D}{d\tau}\,e^{\mu}{}_{\hat \alpha} = \Phi_{\hat \alpha \hat \beta}(\tau)\,e^{\mu \hat \beta}\,.
\end{equation} 
Similarly, let us consider the corresponding world line in a conformally related spacetime with metric $d\tilde{s}^2 = \tilde{g}_{\mu \nu} (x)\,dx^\mu\,dx^\nu$ with proper time $\tilde{\tau}$ and orthonormal tetrad
\begin{equation}\label{V3}
\tilde{e}^{\mu}{}_{\hat \alpha} = \Omega^{-1}(x)\,e^{\mu}{}_{\hat \alpha}\,.
\end{equation} 
The corresponding acceleration tensor is given by
\begin{equation}\label{V4}
\frac{\tilde{D}}{d\tilde{\tau}}\,\tilde{e}^{\mu}{}_{\hat \alpha} = \tilde{\Phi}_{\hat \alpha \hat \beta}(\tilde{\tau})\,\tilde{e}^{\mu \hat \beta}\,,
\end{equation} 
where the covariant differentiation here involves the Christoffel symbols of the conformally related spacetime, namely, 
\begin{equation}\label{V5}
\tilde{\Gamma}^{\alpha}_{\beta \gamma} = \Gamma^{\alpha}_{\beta \gamma} +\left(\frac{\Omega_{,\beta}}{\Omega}\,\delta^\alpha_\gamma + \frac{\Omega_{,\gamma}}{\Omega}\,\delta^\alpha_\beta - \frac{\Omega_{,\delta}}{\Omega}\,g^{\alpha \delta}\, g_{\beta \gamma}\right)\,.
\end{equation} 
From Eqs.~\eqref{V2}--\eqref{V5}, we find, 
\begin{equation}\label{V6}
\tilde{\Phi}_{\hat \alpha \hat \beta} = \frac{1}{\Omega} \left[ \Phi_{\hat \alpha \hat \beta} + \frac{\Omega_{,\mu}}{\Omega}\,(e^{\mu}{}_{\hat \alpha}\,\eta_{\hat \beta \hat 0} - e^{\mu}{}_{\hat \beta}\,\eta_{\hat \alpha \hat 0})\right]\,.
\end{equation} 
This result is a generalization of Eq.~\eqref{U22} and indicates that under conformal rescaling, apart from a scale factor of $\Omega^{-1}$, there is only a contribution to the translational acceleration of the observer and there is no contribution to the rotational acceleration; that is, 
\begin{equation}\label{V7}
\tilde{\Phi}_{\hat 0 \hat i} = \frac{1}{\Omega} \left[ \Phi_{\hat 0 \hat i} + \frac{\Omega_{,\mu}}{\Omega}\,e^{\mu}{}_{\hat i}\right]\,, \qquad \tilde{\Phi}_{\hat i \hat j} = \frac{1}{\Omega}\,\Phi_{\hat i \hat j}\,. 
\end{equation} 
In particular, if one world line is a geodesic, the conformally related world line is accelerated. 

Apropos of the acceleration transformation, we can view Eq.~\eqref{U6} as connecting a Minkowski spacetime with metric $\eta_{\alpha \beta}\, dx'^\alpha \, dx'^\beta$ to another conformally related Minkowski spacetime with metric $f^{-2}(x')\,\eta_{\alpha \beta}\, dx'^\alpha \, dx'^\beta$. That is, in the context of observers in two conformally related Minkowski spacetimes, the inertial observers that are spatially at rest with constant $x'^{i}$ in a global inertial frame of reference correspond to uniformly accelerated observers in the conformally related Minkowski spacetime. Therefore, in Eq.~\eqref{V6} we have $\Phi_{\hat \alpha \hat \beta} = 0$, $e^{\mu}{}_{\hat \alpha} = \delta^\mu_\alpha$ and $\Omega = 1/f$; in this way, Eq.~\eqref{V6} simply reduces to Eq.~\eqref{U22}. 

\subsection{Conformal Invariance of Maxwell's Equations}

Imagine Faraday fields $F_{\mu \nu}$ and $\tilde{F}_{\mu \nu}$  in the conformally related spacetimes. To establish a connection between them, we assume that $F_{\mu \nu} =\tilde{F}_{\mu \nu}$. Next, we note that 
\begin{equation}\label{V8}
\tilde{\nabla}_{\nu}\, \tilde{F}^{\mu \nu} = \frac{1}{\sqrt{-\tilde{g}}}\,\frac{\partial}{\partial x^\nu}(\sqrt{-\tilde{g}}\,\tilde{F}^{\mu \nu}) =  \frac{1}{\Omega^4}\,\nabla_{\nu}\,F^{\mu \nu}\,. 
\end{equation}
It follows that source-free Maxwell's equations,
\begin{equation}\label{V9}
\nabla_{[\rho}\,F_{\mu \nu]} = 0\,, \qquad \nabla_{\nu}\,F^{\mu \nu} = 0\,,
\end{equation}
are conformally invariant. This is a natural consequence of the absence of any intrinsic length scale in source-free electrodynamics~\cite{HeOb}.   

It follows from the conformal invariance of Maxwell's equations and Eq.~\eqref{U6} that Maxwell's equations are invariant under the acceleration transformation. The same is true under constant scaling of Minkowski spacetime coordinates ($x^\mu \mapsto \sigma\,x^\mu$). These results imply that  Maxwell's equations are invariant under the 15-parameter conformal group of Minkowski spacetime. On the other hand, invariant length scales are associated with accelerated observers in relativity theory and hence electromagnetic fields \emph{measured} along conformally related world lines,
\begin{equation}\label{V10}
F_{\hat \alpha \hat \beta} = F_{\mu \nu}\,e^{\mu}{}_{\hat \alpha}\,e^{\nu}{}_{\hat \beta}\,, \qquad  \tilde{F}_{\hat \alpha \hat \beta} = \tilde{F}_{\mu \nu}\,\tilde{e}^{\mu}{}_{\hat \alpha}\,\tilde{e}^{\nu}{}_{\hat \beta}\,, 
\end{equation}
are related by 
\begin{equation}\label{V11}
\tilde{F}_{\hat \alpha \hat \beta} = \Omega^{-2}\,F_{\hat \alpha \hat \beta}\,. 
\end{equation}
Furthermore, for a test particle of mass $M$, charge $Q$ and 4-velocity $u^\mu = dx^\mu/d\tau$, the Lorentz force law
\begin{equation}\label{V12}
\frac{Du^\mu}{d\tau} =  \frac{Q}{M} F^{\mu}{}_\nu\,u^\nu\,, 
\end{equation}
does not remain invariant under conformal rescaling. In fact, 
\begin{equation}\label{V13}
\frac{\tilde{D}\tilde{u}^\mu}{d\tilde{\tau}} =  \frac{Q}{M} \tilde{F}^{\mu}{}_\nu\,\tilde{u}^\nu\, 
\end{equation}
is equivalent to
\begin{equation}\label{V14}
\Omega\,\frac{Du^\mu}{d\tau}+\Omega_{,\rho}\,(u^\rho u^\mu + g^{\rho \mu}) =  \frac{Q}{M} F^{\mu}{}_\nu\,u^\nu\,, 
\end{equation}
which is consistent with $u^\mu\,u_\mu = -1$. 
Thus particle acceleration breaks conformal invariance.  

\subsection{Curvature and Torsion}

In nonlocal gravity~\cite{Mash3c}, general relativity is extended such that one deals with one spacetime metric and two metric-compatible connections. The  Levi-Civita connection is given by the symmetric Christoffel symbols, while the Weitzenb\"ock connection is defined by
\begin{equation}\label{V15}
\Gamma'^{\mu}_{\nu \rho} = E^{\mu}{}_{\hat \alpha} \,\partial_{\nu}\,E_{\rho}{}^{\hat \alpha}\,,
\end{equation}
where $E^{\mu}{}_{\hat \alpha}$ is a smooth tetrad frame field defined on the spacetime manifold. In the conformally related spacetime manifold, the tetrad frame field transforms as in Eq.~\eqref{V3}, so that for the Weitzenb\"ock connection we have  
\begin{equation}\label{V16}
\tilde{\Gamma}'^{\mu}_{\nu \rho} = \Gamma'^{\mu}_{\nu \rho} + \frac{\Omega_{,\nu}}{\Omega}\,\delta^\mu_\rho\,.
\end{equation}
The Levi-Civita connection is torsion free, but has curvature. The transformation of curvature under conformal transformation~\eqref{V1} is well known~\cite{HE, R1}; in particular, the totally traceless Weyl curvature tensor turns out to be conformally invariant, i.e. 
\begin{equation}\label{V17}
\tilde{C}^{\mu}{}_{\nu \rho \sigma} =  C^{\mu}{}_{\nu \rho \sigma}\,. 
\end{equation}

On the other hand, the Weitzenb\"ock connection is curvature free, but has torsion. The tetrad frame field $E^{\mu}{}_{\hat \alpha}$ provides a global network of parallel tetrad frames via the Weitzenb\"ock connection. Using Eq.~\eqref{V16}, it is possible to show that curvature is left invariant under this transformation, namely, $\tilde{R}'^{\mu}{}_{\nu \rho \sigma} =  R'^{\mu}{}_{\nu \rho \sigma}$; in particular, the curvature of the conformally related Weitzenb\"ock connection vanishes as well. Therefore, the notion of a teleparallel frame field is in this sense conformally invariant. Moreover, for the torsion tensor
\begin{equation}\label{V18}
C_{\mu \nu}{}^{\alpha} = \Gamma'^{\alpha}_{\mu \nu} - \Gamma'^{\alpha}_{\nu \mu}\,,
\end{equation}
we find  
\begin{equation}\label{V19}
\tilde{C}_{\mu \nu}{}^{\alpha} = C_{\mu \nu}{}^{\alpha} +  \frac{\Omega_{,\mu}}{\Omega}\,\delta^\alpha_\nu -  \frac{\Omega_{,\nu}}{\Omega}\,\delta^\alpha_\mu\,.
\end{equation}
Similarly, for the contorsion tensor,
\begin{equation}\label{V20}
K_{\mu \nu}{}^{\alpha} = \Gamma^{\alpha}_{\mu \nu} - \Gamma'^{\alpha}_{\mu \nu}\,,
\end{equation}
we find
\begin{equation}\label{V21}
\tilde{K}_{\mu \nu}{}^{\alpha} = K_{\mu \nu}{}^{\alpha} +  \frac{\Omega_{,\nu}}{\Omega}\,\delta^\alpha_\mu -  \frac{\Omega_{,\rho}}{\Omega}\,g^{\alpha \rho}\,g_{\mu \nu}\,.
\end{equation}

It is interesting to define a new \emph{conformal torsion tensor} by 
\begin{equation}\label{V22}
\mathbb{C}^{\alpha}{}_{\beta \gamma} = g^{\alpha \delta}\left[C_\delta\, g_{\beta \gamma} - C_\beta\, g_{\delta \gamma} + \frac{1}{2}\,(C_{\gamma \beta \delta} -  C_{\gamma \delta \beta}) +\frac{5}{2} \,C_{\delta \beta \gamma}\right]\,,
\end{equation}
where $C_\mu$ is the \emph{torsion vector} given by the trace of the torsion tensor, 
\begin{equation}\label{V23}
C_\mu = - C_{\mu}{}^{\nu}{}_{\nu} = C^{\alpha}{}_{\mu \alpha}\,. 
\end{equation}
It is straightforward to check that $\mathbb{C}_{\alpha \beta \gamma}$ is antisymmetric in its first two indices and is totally traceless. Moreover, in the conformally related spacetime, we have
\begin{equation}\label{V24}
\tilde{C}_\mu =C_{\mu}  - 3\, \frac{\Omega_{,\mu}}{\Omega}\,. 
\end{equation}
Using Eqs.~\eqref{V19} and~\eqref{V24}, we find that the conformal torsion tensor is indeed conformally invariant; that is,
\begin{equation}\label{V25}
\tilde{\mathbb{C}}^{\alpha}{}_{\beta \gamma} = \mathbb{C}^{\alpha}{}_{\beta \gamma}\,.
\end{equation}
The analogy between Eqs.~\eqref{V17} and~\eqref{V25} is noteworthy. The completely traceless and conformally invariant torsion tensor $\mathbb{C}^{\mu}{}_{\nu \rho}$ is the analog of the completely traceless and conformally invariant Weyl curvature tensor $C^{\mu}{}_{\nu \rho \sigma}$.

\section{Discussion}

We have studied a specific congruence of accelerated observers in Minkowski spacetime. The corresponding accelerated system is related to the conformal group of Minkowski spacetime. All actual observers are accelerated. One can construct intrinsic acceleration scales from the acceleration of the observer and the speed of light. The presence of an acceleration scale breaks conformal invariance. This important point has been illustrated in detail in this paper.  

A possible extension of our treatment would involve, for instance, the formulation of nonlocal electrodynamics for the accelerated observers investigated in this work.

\section*{ACKNOWLEDGMENTS}

I am grateful to Friedrich Hehl for his helpful comments on the manuscript.



\end{document}